\documentclass[aps,prl,twocolumn,groupedaddress]{revtex4}

\usepackage{graphicx}

\begin{document}


\title{Negative Quantum Capacitance of Carbon Nanotube Field-Effect Transistors}


\author{L. Latessa, A. Pecchia, A. Di Carlo}
\affiliation{INFM - University of Rome ``Tor Vergata'', Via del Politecnico 1, 00133, Roma, Italy}
\author{P. Lugli}
\affiliation{Lehrstuhl f\"ur Nanoelektronik, TU M\"unchen, Arcistrasse 21, D-80333, M\"unchen, Germany}

\date{\today}

\begin{abstract}
Atomistic density functional theory (DFT) calculations of the capacitance between a metallic cylindric gate and a carbon nanotube (CNT) are reported. Results stressing the predominant effect of quantum capacitance in limiting or even enhancing screening properties of the CNT are shown. Other contributions to the quantum capacitance beyond the electronic density of state (DOS) are pointed out. Negative values of the quantum capacitance are obtained for low-density systems, which correspondingly over-screen the gate field. This unconventional behavior of the quantum capacitance is related to the predominance of the exchange contribution in the total electronic energy of the CNT.   
\end{abstract}


\maketitle


Theoretical investigation of the physical properties of realistic quasi one-dimensional electronic systems like CNTs becomes more and more attractive due to the promising perspectives they offer to nanotecnology.
Especially interesting are the electronic properties of a CNT when it is used as a quasi-one dimensional channel in nanoscaled field-effect devices \cite{Appenzeller04,Leonard,Lundstrom}. 
The reduced dimensionality of these systems offers the possibility of directly investigating the fingerprints of the many-body nature of the electron-electron interactions in the screening phenomena.
In this Letter we report on atomistic DFT calculations of the screening properties of a coaxially gated CNT (10,0). 
The signature of many-body physics manifests itself at low carrier densities where the CNT (10,0) exhibits a singular screening behavior, turning from a poorly screening electron gas to a classical metal and eventually resulting into an over-screening situation.\\\indent
It is known that for a system of low-dimensional conductors the classical electrostatic concept of capacitance can no longer be applied. 
Luryi reported \cite{Luryi} that a two-dimensional electron gas (2DEG) used as grounded middle plate in a three plates capacitor is not able to screen the field emanating from the charges on one of the other plates; on the contrary, the field partially penetrates through the 2DEG and induces charge on the opposite external plate. 
This non-classical field penetration is due to the reduced screening properties of the 2DEG, and it changes the capacitance from a geometrical quantity to an electrochemical one \cite{Buttiker}. 
In a circuital approach, this means that new quantum capacitances will appear besides the geometrical ones to describe the non-classical (quantum) screening properties of such low-dimensional electronic systems \cite{Luryi}.\\\indent
Quantum capacitance is however a more general concept which is strictly connected to that of thermodynamic compressibility. 
If we refer to an interacting one-dimensional electronic gas, the compressibility $K$ is defined as \cite{Fetter} $K^{-1}=n^{2}d^{2}E_{tot}/dn^{2}=n^{2}d\mu/dn$, where $E_{tot}$, $\mu$, and $n$ are, respectively, the total ground-state energy (per unit length), the chemical potential and the electronic density of the system. 
The quantum capacitance per unit length $C_{Q}$ is defined as $1/C_{Q}=(1/e^{2})d\mu /dn$ \cite{Smith}. Clearly the capacitance results to be proportional to the compressibility. 
Combining the expressions for $C_{Q}$ and for $K$ one also finds $L/C_{Q}=1/e^{2}\rho_{0}(\epsilon_{F})\cdot K_{0}/K$, where $\rho_{0}(\epsilon_{F})$ and $K_{0}$ are, respectively, the DOS at the Fermi energy $\epsilon_{F}$ and the compressibility of the free electron gas.
This result shows that neglecting electron-electron interaction ($K_{0}/K=1$), the quantum capacitance can be simply obtained from the DOS at the Fermi level. 
An exact evaluation of quantum capacitance is however of fundamental importance for nanoelectronic devices such as CNT field-effect transistors (CNTFET). 
Gate modulation depends on the combined effect of the insulator and quantum capacitance. Up to now, however, calculations on CNTFETs have only considered the proportionality between $C_{Q}$ and the DOS \cite{Rahman} neglecting electron-electron interactions effects, which may induce strong variations of $C_{Q}$. Model Hamiltonian approaches show that, due to the exchange interactions, the compressibility of quasi one- and two-dimensional electronic systems may also be negative \cite{Calmels96}.
A negative sign of the compressibility has also been found in capacitive measurements on 2DEG realized in GaAs quantum wells \cite{Eisenstein}. 
The purpose of this work is to shed some light on the quantum capacitance issue for CNTs, clarifying its dependence on electron-electron interactions. 
We demonstrate that the quantum capacitance of a CNT is not just determined by the DOS. 
\emph{In other words, the quasi one-dimensional nature of a CNT, reflected in a small DOS, may not only result in a poor screening behavior. On the contrary, when the carrier density is sufficiently low, due to the exchange interactions the CNT over-screens the external gate field and its quantum capacitance assumes negative values}.

\begin{figure}
\includegraphics[width=6.9cm,angle=0]{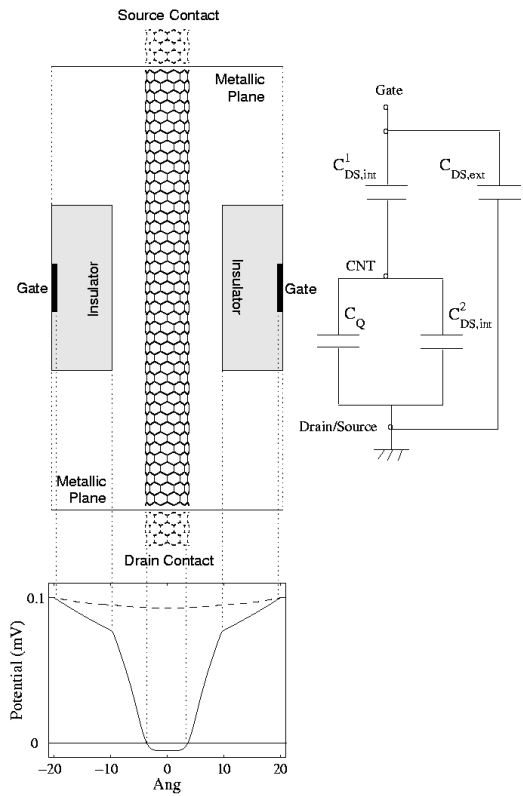}
\caption{Coaxially gated (10,0) semiconducting CNT. The bare and screened electrostatic potential are shown as a dashed and solid line respectively. Plots refer to the cross-section taken in the middle of the gate contact. Potential inside the CNT is clearly over-screened.\label{fig:system}}
\end{figure}
In Fig. \ref{fig:system} the schematic section of the simulated coaxially gated CNT (10,0) is shown.
The self-consistent screened potential needed for capacitance calculations is obtained solving in real space a three dimensional Poisson equation. The metallic gate contact is treated as an appropriate boundary condition for the electrostatic problem. 
The gate is surrounded by an insulator layer with dielectric constant $\epsilon_{r} = 3.9$, which prevents the CNT to touch the metal in realistic field-effect devices. 
Charge injection inside the CNT is obtained using two semi-infinite portions of CNT (10,0) which act as source and drain reservoirs \cite{Book}.
Dirichlet boundary conditions have been used on the Poisson calculation box correspondingly to the source and drain contacts; this introduces two metallic surfaces in the system (see Fig. \ref{fig:system}).\\\indent
Calculations have been performed using the non-equilibrium Green's function \cite{non_eq} density functional tight-binding (gDFTB \cite{Review,Book}) approach, which is based on a local density approximation (LDA) of the exchange-correlation (XC) functional. The XC contribution to the electronic energy is modeled as an on-site Hubbard energy \cite{Elstner}. 
It is important to note that the value of this Hubbard parameter takes into account the attraction induced by the exchange contribution to the on-site energy, and results to be smaller than an on-site parameter which just corresponds to the Hartree repulsion energy.
Calculations have been performed using only $p_{z}$ orbitals for carbon atoms. This allow us to simulate a 21.3 nm long CNT (10,0) (corresponding to 50 unit cells). We stress that we obtain the same over-screening results simulating a shorter CNT modeled with a complete $sp^{3}$ basis. 
To obtain density dependent results, we artificially dope the semiconducting CNT (10,0) varying the number of valence electrons per carbon atom. In this doping model any added extra charge is uniformly distributed over the whole CNT, leaving the band structure unaffected \cite{Tersoff}.\\\indent
Several theoretical methods have been proposed in the literature to extract the compressibility of a low dimensional electron gas. Standard approaches consist on numerical evaluations of the second derivative of the total energy with respect to the density \cite{Calmels93,Eisenstein}. 
Other methods based on the evaluation of a static structure factor which takes into account finite width effects in two and one-dimensional electron gases have also been proposed \cite{Calmels96}. 
In this work we use a capacitive model (see Fig. \ref{fig:system}) to directly extract the sign and the magnitude of the quantum capacitance at an atomistic scale. 
A different atomistic approach based on the evaluation of a properly defined inductance matrix \cite{Buttiker} has been also presented \cite{HGuo}.\\\indent  
The circuital model shown in Fig. \ref{fig:system} extends the equivalent circuit used in Ref. \cite{Luryi} to a one-dimensional system with a locally applied gate field.
The series connection of the geometrical capacitances  $C_{\scriptscriptstyle DS,int}^{1}$ and $C_{\scriptscriptstyle DS,int}^{2}$ describes the capacitive induction between the gate and those parts of the source(drain) metallic surfaces internal to the CNT. 
Using two capacitances we can account for the fact that this capacitive induction is not direct, but it is mediated by the CNT.
In other words, part of the charge that the gate field would induce on the inner parts of the source(drain) metallic surfaces is induced on the CNT. 
The quantum capacitance $C_{Q}$ just accounts for the capability of the CNT of localizing this charge on its surface. 
When there are no mobile electrons able to screen the gate field, we have $C_{Q}=0$ and charge is induced only on the source(drain) metallic surfaces. 
On the other hand, in the classical limit ($|C_{Q}| \rightarrow \infty$) the CNT completely screens the gate field, hence no charge is induced on the source(drain) metallic surfaces inside the CNT itself. 
Finite values of $C_{Q}$ describe partial induction of charge on the tube. 
The capacitance $C_{\scriptscriptstyle DS,ext}$ accounts for the charge induced by the gate field on those parts of the metallic source(drain) surfaces external to the CNT. 
If the length of the CNT is large with respect to both the gate length and the insulator thickness, the charge induced on the metallic source(drain) surfaces is small with respect to the charge induced on the CNT. 
In this situation we can neglect both $C_{\scriptscriptstyle DS,int}^{2}$ and $C_{\scriptscriptstyle DS,ext}$.
This approximation apply to our system, where $C_{\scriptscriptstyle DS,int}^{2}/|C_{Q}|\lesssim 10^{-4}$ and $C_{\scriptscriptstyle DS,ext}$ is at least two orders of magnitude smaller than the series of $C_{\scriptscriptstyle DS,int}^{1}$ and $C_{Q}$ in the explored range of carrier densities \cite{Field}. 
Starting from the simplified circuital model, we obtain the following meaningful expression for $C_{Q}$
\begin{equation} 
\label{eq:CQ}
C_{Q}\left(n\right) = C_{\scriptscriptstyle DS,int}^{1}\left(\frac{\delta Q\left(n_{\scriptscriptstyle TS}\right)}
                                                    {\delta Q\left(n\right)}-1\right)^{-1},
\end{equation}
where $\delta Q$ is the charge induced on the CNT by a small applied gate bias $\delta V_{G}$, $n$ is the CNT carrier density and $n_{\scriptscriptstyle TS}$ is the carrier density value which causes the CNT to totally screen (TS) the gate field. 
Clearly a negative quantum capacitance is obtained at those carrier densities for which the CNT is able to localize more charge than what strictly needed to completely screen the gate field. 
We note that $C_{\scriptscriptstyle DS,int}^{1}$ is just the geometric insulator capacitance of the cylindric capacitor. Nevertheless, due to the strong fringing field effects it can not be analytically evaluated. We operatively compute it as $\delta Q\left(n_{\scriptscriptstyle TS}\right)/\delta V_{G}$, referring to the limit situation in which the CNT completely screens the gate field and a classical cylindric capacitor is recovered.

\begin{figure}
\includegraphics[width=6.8cm,angle=270]{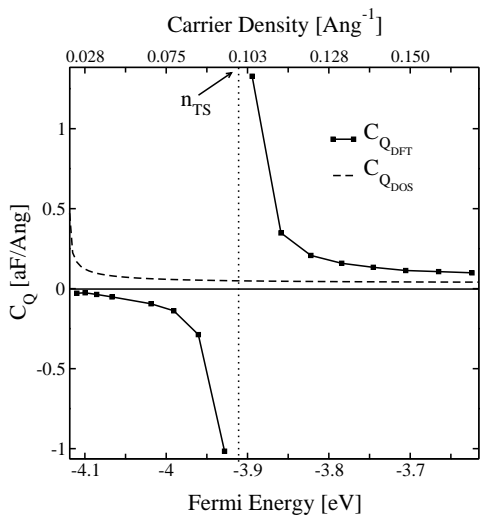}
\caption{Computed quantum capacitance per unit length of the CNT (10,0) as a function of both the Fermi level and the carrier density. Only the first subband of the CNT is filled. Solid line refers to DFT calculations. Dashed line is obtained just considering the DOS contribution to $C_{Q}$.\label{fig:Cq}}
\end{figure}

In Fig. \ref{fig:system} we report the cross-section of the screened and bare potentials taken in the middle of the gate contact. 
Both bare and screened potentials correspond to a gate bias $\delta V_{G}=10^{-4}V$ and a bias of $0$ $V$ applied to both the source and drain metallic contacts. 
The CNT (10,0) is lightly $n$-doped, resulting in a small carrier density $n=4.7 \cdot 10^{-2} \AA^{-1}$. 
In this situation the CNT (10,0) over-screens the gate field, as clearly shown by the presence of an over-screened negative electrostatic potential inside the CNT itself. 
We stress that the over-screening result does not depend on the particular doping model we have used. In fact, a negative quantum capacitance has been also obtained for an intrinsic metallic CNT (5,0) \cite{Cabria}. 
The only advantage in using a $n$-doped semiconducting CNT is that we can continuously vary the carrier densities inside the first conduction subband.\\\indent
The presence of an additional contribution to the CNT quantum capacitance beyond the DOS can be appreciated in Fig. \ref{fig:Cq}. Here we plot the quantum capacitance computed for several carrier density values which correspond to a given Fermi level position inside the first CNT subband. 
Solid line refers to DFT calculations ($C_{Q_{DFT}}$), while the dashed line is just obtained plotting $C_{Q_{DOS}}=e^{2}\rho(\epsilon_{F})$, where $\rho(\epsilon_{F})$ is the DOS of the CNT computed at the Fermi level. 
$C_{Q_{DOS}}$ is the limit which is recovered by the DFT calculation for large enough carrier densities. As the density lowers, $C_{Q_{DFT}}$ tends to an infinite positive value. 
This singular behavior corresponds to the critical density $n_{\scriptscriptstyle TS}$ for which the CNT completely screens the external field. Below this critical density the CNT accumulates more charge than what strictly needed for total screening, and an over-screening situation is generated. 
We note that a negative value of the quantum capacitance does not mean necessarily instability. The presence of a positive insulator capacitance with a smaller absolute value than $C_{Q}$ gives a positive value to the total capacitance, thus stabilizing the whole system \cite{Eisenstein}.

\begin{figure}
\includegraphics[width=6.0cm,angle=270]{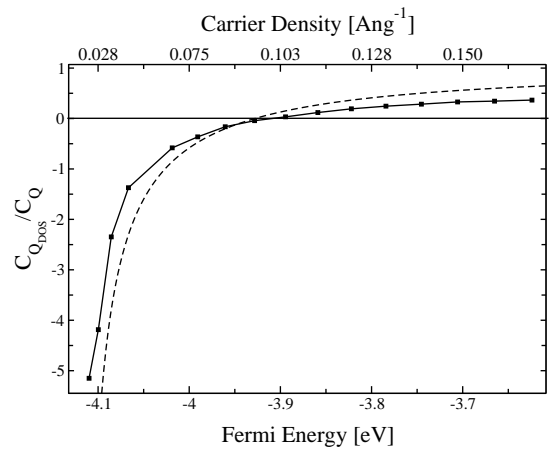}
\caption{Electron-electron interaction contribution to the quantum capacitance of a CNT (10,0). Solid line refers to the DFT atomistic calculation. Dashed line is an analytical result for the ratio $C_{Q_{DOS}}/C_{Q_{HF}}$ obtained for a parabolic-confined quantum wire matching the CNT.\label{fig:ratio}}
\end{figure}

The presence of a negative quantum capacitance is related to the attractive exchange energy contribution, which at low carrier densities becomes predominant producing a local minimum in the total energy and a change in the sign of its second derivative (i.e. in the sign of the compressibility $K$) \cite{Fetter}.
This theoretical prediction is confirmed by results we obtain eliminating the XC energy contribution in the Hubbard on-site parameter: in this case our calculations result in a positive screened potential inside the CNT, on which in fact a smaller charge than $\delta Q\left(n_{\scriptscriptstyle TS}\right)$ is induced.\\\indent 
As explained earlier, the ratio $C_{Q_{DOS}}/C_{Q_{DFT}}$ is representative of the ratio $K_{0}/K_{DFT}$, where $K_{DFT}$ is the DFT compressibility of the CNT. 
Analytical evaluations of the effects that electron-electron interactions have on the compressibility of various quantum wire models have been presented in the literature \cite{Calmels96,Calmels95_Corr}. 
We observe that if the radius of the CNT we deal with is not too big respect to the orbital extension of the carbon atom, the CNT can be suitably approximated by a parabolic confined cylindrical quantum wire. 
A negative compressibility regime has been observed in such systems in the Hartree-Fock (HF) approximation, where electron-electron interactions are included at the lowest order as an exchange energy.
Using these analytical results we have evaluated the ratio $C_{Q_{DOS}}/C_{Q_{HF}}$, obtaining
\begin{equation}
\label{eq:ratioK}
\frac{C_{Q_{DOS}}}{C_{Q_{HF}}}=1-\frac{1}
                       {\displaystyle 2\pi a^{*}\sqrt{\frac{2m^{*}E_{F}}{\hbar^{2}}}}
                f\left(2b\sqrt{\frac{2m^{*}E_{F}}{\hbar^{2}}}\right).
\end{equation}
The function $f$ is the static form factor which takes into account width effects associated to the quasi one-dimensional nature of the electron gas and $b$ is the width parameter of the quantum wire. 
The quantities $E_{F}$ and $m^{*}$ are respectively the Fermi level and the effective mass of the electronic system, while $a^{*}=\hbar^{2}/m^{*}e^{2}$ is the effective Bohr radius defined with $m^{*}$. 
Results obtained using a width parameter $b=4 \AA$ and an effective electron mass $m^{*} \simeq 0.083$ $m_{e}$ (matching respectively the effective width and the first subband effective mass of the CNT (10,0)) are shown in Fig. \ref{fig:ratio} together with the ratio $C_{Q_{DOS}}/C_{Q_{DFT}}$ obtained within our approach. 
Comparing our atomistic DFT results with the analytical calculations, it is important to observe that in our model the exchange energy is only included in a local approximation, and that the Hubbard approximation to the XC functional takes also into account correlation energy contributions. Moreover, our atomistic model includes the Hartree interaction which is completely absent in a uniform electron gas model due to the presence of a neutralizing positive background. Respect to purely HF models, the correlation contribution decreases the inverse compressibility, while the direct Coulomb contribution increases it \cite{Calmels95_Corr}.
Nevertheless, the good qualitative agreement of results presented in Fig. \ref{fig:ratio} shows that the fundamental relationship between electron-electron interactions and over-screening in a CNT is correctly predicted by both DFT and HF calculations.\\\indent
Results shown both in Fig. \ref{fig:ratio} and in Fig. \ref{fig:Cq} refer to a range of densities for which only the first subband is occupied. Adding further carriers, the second subband of the CNT starts to be filled and a more complex behavior of the quantum capacitance appears (namely, we observe again a transition to negative values when the second subband starts to be filled).\\\indent  
To summarize, atomistic DFT simulations have revealed that a CNT can over-screen an external field produced by a gate contact. Negative values of the CNT quantum capacitance have been obtained in a low density regime inside the first conduction subband. This demonstrate that the DOS is not the only contribution to the capacitance of a gated CNT, and that electron-electron interactions effects can assume a fundamental role in determining the screening properties of a gated nanoscale device and consequently its field-effect dependence.


\begin{thebibliography}{20}
\expandafter\ifx\csname natexlab\endcsname\relax\def\natexlab#1{#1}\fi
\expandafter\ifx\csname bibnamefont\endcsname\relax
  \def\bibnamefont#1{#1}\fi
\expandafter\ifx\csname bibfnamefont\endcsname\relax
  \def\bibfnamefont#1{#1}\fi
\expandafter\ifx\csname citenamefont\endcsname\relax
  \def\citenamefont#1{#1}\fi
\expandafter\ifx\csname url\endcsname\relax
  \def\url#1{\texttt{#1}}\fi
\expandafter\ifx\csname urlprefix\endcsname\relax\def\urlprefix{URL }\fi
\providecommand{\bibinfo}[2]{#2}
\providecommand{\eprint}[2][]{\url{#2}}

\bibitem[{\citenamefont{Appenzeller et~al.}(2004)\citenamefont{Appenzeller,
  Knoch, Derycke, Martel, Wind, and Avouris}}]{Appenzeller04}
\bibinfo{author}{\bibfnamefont{J.}~\bibnamefont{Appenzeller}},
  \bibinfo{author}{\bibfnamefont{J.}~\bibnamefont{Knoch}},
  \bibinfo{author}{\bibfnamefont{V.}~\bibnamefont{Derycke}},
  \bibinfo{author}{\bibfnamefont{R.}~\bibnamefont{Martel}},
  \bibinfo{author}{\bibfnamefont{S.}~\bibnamefont{Wind}}, \bibnamefont{and}
  \bibinfo{author}{\bibfnamefont{P.}~\bibnamefont{Avouris}},
  \bibinfo{journal}{Rep.\ Prog.\ Phys.} \textbf{\bibinfo{volume}{67}},
  \bibinfo{pages}{1} (\bibinfo{year}{2004}).

\bibitem[{\citenamefont{Leonard and Tersoff}(2002)}]{Leonard}
\bibinfo{author}{\bibfnamefont{F.}~\bibnamefont{Leonard}} \bibnamefont{and}
  \bibinfo{author}{\bibfnamefont{J.}~\bibnamefont{Tersoff}},
  \bibinfo{journal}{Phys.\ Rev.\ Lett.} \textbf{\bibinfo{volume}{88}},
  \bibinfo{pages}{258302} (\bibinfo{year}{2002}).

\bibitem[{\citenamefont{Guo et~al.}(2002)\citenamefont{Guo, Lundstrom, and
  Datta}}]{Lundstrom}
\bibinfo{author}{\bibfnamefont{J.}~\bibnamefont{Guo}},
  \bibinfo{author}{\bibfnamefont{M.}~\bibnamefont{Lundstrom}},
  \bibnamefont{and} \bibinfo{author}{\bibfnamefont{S.}~\bibnamefont{Datta}},
  \bibinfo{journal}{Appl.\ Phys.\ Lett.} \textbf{\bibinfo{volume}{80}},
  \bibinfo{pages}{3192} (\bibinfo{year}{2002}).

\bibitem[{\citenamefont{Luryi}(1988)}]{Luryi}
\bibinfo{author}{\bibfnamefont{S.}~\bibnamefont{Luryi}},
  \bibinfo{journal}{Appl.\ Phys.\ Lett.} \textbf{\bibinfo{volume}{52}},
  \bibinfo{pages}{501} (\bibinfo{year}{1988}).

\bibitem[{\citenamefont{B{\"u}ttiker}(1993)}]{Buttiker}
\bibinfo{author}{\bibfnamefont{M.}~\bibnamefont{B{\"u}ttiker}},
  \bibinfo{journal}{J.\ Phys.:\ Condens.\ Matter} \textbf{\bibinfo{volume}{5}},
  \bibinfo{pages}{9361} (\bibinfo{year}{1993}).

\bibitem[{\citenamefont{Fetter and Walecka}(1971)}]{Fetter}
\bibinfo{author}{\bibfnamefont{A.~L.} \bibnamefont{Fetter}} \bibnamefont{and}
  \bibinfo{author}{\bibfnamefont{J.~D.} \bibnamefont{Walecka}},
  \emph{\bibinfo{title}{Quantum Theory of Many particle Systems}}
  (\bibinfo{publisher}{McGraw-Hill}, \bibinfo{address}{New York},
  \bibinfo{year}{1971}).

\bibitem[{\citenamefont{Smith et~al.}(1986)\citenamefont{Smith, Wang, and
  Stiles}}]{Smith}
\bibinfo{author}{\bibfnamefont{T.~P.} \bibnamefont{Smith}},
  \bibinfo{author}{\bibfnamefont{W.~I.} \bibnamefont{Wang}}, \bibnamefont{and}
  \bibinfo{author}{\bibfnamefont{P.~J.} \bibnamefont{Stiles}},
  \bibinfo{journal}{Phys.\ Rev.\ B} \textbf{\bibinfo{volume}{34}},
  \bibinfo{pages}{R2995} (\bibinfo{year}{1986}).

\bibitem[{\citenamefont{Rahman et~al.}(2003)\citenamefont{Rahman, Guo, Datta,
  and Lundstrom}}]{Rahman}
\bibinfo{author}{\bibfnamefont{A.}~\bibnamefont{Rahman}},
  \bibinfo{author}{\bibfnamefont{J.}~\bibnamefont{Guo}},
  \bibinfo{author}{\bibfnamefont{S.}~\bibnamefont{Datta}}, \bibnamefont{and}
  \bibinfo{author}{\bibfnamefont{M.}~\bibnamefont{Lundstrom}},
  \bibinfo{journal}{IEEE\ Trans.\ Electr.\ Dev.} \textbf{\bibinfo{volume}{50}},
  \bibinfo{pages}{1853} (\bibinfo{year}{2003}).

\bibitem[{\citenamefont{Calmels and Gold}(1996)}]{Calmels96}
\bibinfo{author}{\bibfnamefont{L.}~\bibnamefont{Calmels}} \bibnamefont{and}
  \bibinfo{author}{\bibfnamefont{A.}~\bibnamefont{Gold}},
  \bibinfo{journal}{Phys.\ Rev.\ B} \textbf{\bibinfo{volume}{53}},
  \bibinfo{pages}{10846} (\bibinfo{year}{1996}).

\bibitem[{Eis()}]{Eisenstein}
\bibinfo{note}{J. P. Eisenstein, L. N. Pfeiffer and K. W. West, Phys. Rev.
  Lett. {\bf 68}, 674 (1992); Phys. Rev. B {\bf 50}, 1760 (1994)}.

\bibitem[{\citenamefont{Di~Carlo et~al.}(2005)\citenamefont{Di~Carlo, Pecchia,
  Latessa, Fraunheim, and Seifert}}]{Book}
\bibinfo{author}{\bibfnamefont{A.}~\bibnamefont{Di~Carlo}},
  \bibinfo{author}{\bibfnamefont{A.}~\bibnamefont{Pecchia}},
  \bibinfo{author}{\bibfnamefont{L.}~\bibnamefont{Latessa}},
  \bibinfo{author}{\bibfnamefont{T.}~\bibnamefont{Fraunheim}},
  \bibnamefont{and} \bibinfo{author}{\bibfnamefont{G.}~\bibnamefont{Seifert}},
  \emph{\bibinfo{title}{Introducing molecular electronics}}
  (\bibinfo{publisher}{Springer-Verlag}, \bibinfo{address}{New York},
  \bibinfo{year}{2005}).

\bibitem[{non()}]{non_eq}
\bibinfo{note}{The non-equilibrium Green's function (NEGF) formalism is usually
  used for quantum transport computations. Although no source/drain bias is
  applied in our calculations, the NEGF approach allows us to properly treat
  open boundary conditions for our system. The charge density and the
  electrostatic potential we obtain are therefore determined by the scattering
  states of an infinite CNT and no finite-length effects are introduced.}

\bibitem[{\citenamefont{Pecchia and Di~Carlo}(2004)}]{Review}
\bibinfo{author}{\bibfnamefont{A.}~\bibnamefont{Pecchia}} \bibnamefont{and}
  \bibinfo{author}{\bibfnamefont{A.}~\bibnamefont{Di~Carlo}},
  \bibinfo{journal}{Rep.\ Prog.\ Phys.} \textbf{\bibinfo{volume}{67}},
  \bibinfo{pages}{1} (\bibinfo{year}{2004}).

\bibitem[{\citenamefont{Elstner et~al.}(1998)\citenamefont{Elstner, Porezag,
  Jungnickel, Elsner, Haugk, Frauenheim, Suhai, and Seifert}}]{Elstner}
\bibinfo{author}{\bibfnamefont{M.}~\bibnamefont{Elstner}},
  \bibinfo{author}{\bibfnamefont{D.}~\bibnamefont{Porezag}},
  \bibinfo{author}{\bibfnamefont{G.}~\bibnamefont{Jungnickel}},
  \bibinfo{author}{\bibfnamefont{J.}~\bibnamefont{Elsner}},
  \bibinfo{author}{\bibfnamefont{M.}~\bibnamefont{Haugk}},
  \bibinfo{author}{\bibfnamefont{T.}~\bibnamefont{Frauenheim}},
  \bibinfo{author}{\bibfnamefont{S.}~\bibnamefont{Suhai}}, \bibnamefont{and}
  \bibinfo{author}{\bibfnamefont{G.}~\bibnamefont{Seifert}},
  \bibinfo{journal}{Phys.\ Rev.\ B} \textbf{\bibinfo{volume}{58}},
  \bibinfo{pages}{7260} (\bibinfo{year}{1998}).

\bibitem[{\citenamefont{Leonard and Tersoff}(1999)}]{Tersoff}
\bibinfo{author}{\bibfnamefont{F.}~\bibnamefont{Leonard}} \bibnamefont{and}
  \bibinfo{author}{\bibfnamefont{J.}~\bibnamefont{Tersoff}},
  \bibinfo{journal}{Phys.\ Rev.\ Lett.} \textbf{\bibinfo{volume}{83}},
  \bibinfo{pages}{5174} (\bibinfo{year}{1999}).

\bibitem[{\citenamefont{Gold and Calmels}(1993)}]{Calmels93}
\bibinfo{author}{\bibfnamefont{A.}~\bibnamefont{Gold}} \bibnamefont{and}
  \bibinfo{author}{\bibfnamefont{L.}~\bibnamefont{Calmels}},
  \bibinfo{journal}{Phys.\ Rev.\ B} \textbf{\bibinfo{volume}{48}},
  \bibinfo{pages}{11622} (\bibinfo{year}{1993}).

\bibitem[{\citenamefont{Pomorski et~al.}(2004)\citenamefont{Pomorski, Pastewka,
  Roland, Guo, and Wang}}]{HGuo}
\bibinfo{author}{\bibfnamefont{P.}~\bibnamefont{Pomorski}},
  \bibinfo{author}{\bibfnamefont{L.}~\bibnamefont{Pastewka}},
  \bibinfo{author}{\bibfnamefont{C.}~\bibnamefont{Roland}},
  \bibinfo{author}{\bibfnamefont{H.}~\bibnamefont{Guo}}, \bibnamefont{and}
  \bibinfo{author}{\bibfnamefont{J.}~\bibnamefont{Wang}},
  \bibinfo{journal}{Phys.\ Rev.\ B} \textbf{\bibinfo{volume}{69}},
  \bibinfo{pages}{115418} (\bibinfo{year}{2004}).

\bibitem[{Fie()}]{Field}
\bibinfo{note}{The charge induced on the metallic surfaces is evaluated
  starting from the local electric field ($E_{\perp}=\sigma /\epsilon_{0}$).}

\bibitem[{Cab()}]{Cabria}
\bibinfo{note}{The metallic or semiconducting nature predicted in the graphene
  sheet model (a CNT (n,m) is metallic if $mod((n-m)/3)=0)$ can be invalidated
  due to curvature effects in narrow zig-zag CNTs (see for example: J. Cabria,
  J. W. Mintmire and C. T. White, Int. J. Quant. Chem. {\bf 91}, 51 (2003)).}

\bibitem[{\citenamefont{Calmels and Gold}(1995)}]{Calmels95_Corr}
\bibinfo{author}{\bibfnamefont{L.}~\bibnamefont{Calmels}} \bibnamefont{and}
  \bibinfo{author}{\bibfnamefont{A.}~\bibnamefont{Gold}},
  \bibinfo{journal}{Phys.\ Rev.\ B} \textbf{\bibinfo{volume}{52}},
  \bibinfo{pages}{10841} (\bibinfo{year}{1995}).

\end{thebibliography}
\end{document}